\documentclass[aps,prl,floatfix,epsfig,twocolumn,showpacs,preprintnumbers]{revtex4}
\usepackage{amssymb}
\usepackage{graphicx}
\usepackage{amsmath}

\newcommand{\vk}{{\mathbf{k}}}
\renewcommand{\a}{\alpha}
\renewcommand{\b}{\beta}

\begin{document}

\title{The $\alpha \rightarrow \gamma$ transition in Ce: a theoretical
view from optical spectroscopy.}  \author{Kristjan Haule$^{a,d}$,
Viktor Oudovenko$^{b,d}$, Sergej Y. Savrasov$^{c}$, Gabriel
Kotliar$^{d}$} \affiliation{$^{a}$Jo\v zef Stefan Institute, SI-1000
Ljubljana, Slovenia} \affiliation{$^{b}$Laboratory for Theoretical
Physics, Joint Institute for Nuclear Research, 141980 Dubna, Russia}
\affiliation{$^{c}$Department of Physics, New Jersey Institute of
Technology, Newark, NJ 07102, USA} \affiliation{$^{d}$Department of
Physics and Center for Material Theory, Rutgers University, Pscataway,
NJ 08854, USA} \date{\today}

\begin{abstract}
Using a novel approach to calculate optical properties of strongly
correlated systems, we address the old question of the physical origin of
the $\alpha \rightarrow \gamma $ transition in Ce. We find that the Kondo
collapse model describes the optical data better than the Mott transition
picture. Our results compare well with existing thin--film experiments. We
predict the full temperature dependence of the optical spectra and find the
development of a pseudogap around 0.6 eV in the vicinity of the $\alpha
\rightarrow \gamma$ phase transition.
\end{abstract}

\date{\today }
\maketitle

At a temperature of 600~K and pressure less than 20~Kbars, elemental Cerium
undergoes a transition between two isostructural phases, a high pressure
phase or $\alpha$ phase and a low pressure $\gamma$ phase. In $\alpha $--Ce the $f$
electron is delocalized (for example the spin susceptibility is temperature
independent ) while in $\gamma $--Ce the $f$ electron is localized (for
example the electron has a Curie--like susceptibility). Several basic
questions about this transition are still being debated. What is the driving
mechanism of this transition? What is the role of the $spd$ electrons which
are near the Fermi level in this material? To address these questions two
main hypothesis have been advanced. Johansson proposed a Mott transition
scenario \cite{Johansson}, where the transition is connected to
delocalization of the $f$ electron. In the $\alpha $ phase the $f$ electron
is itinerant, band like while in the $\gamma $ phase it is localized and
hence does not participate in the bonding, explaining the volume collapse.
In this picture the $spd$ electrons are mere spectators well described by
the density functional theory calculations (DFT) in the local density
approximation (LDA) or its extensions. The $f$ electrons in $\alpha $--Ce
can also be treated in the band picture while in $\gamma $--Cerium should be
considered as part of the core, and do not participate in the bonding.

A different view on this problem was proposed by Allen and Martin who
introduced the Kondo volume collapse model for the $\alpha -\gamma $
transition. They suggested that the transition was connected with
modifications in the effective hybridization of the $spd$ bands with
the $f$-electron.  In this picture the main change when going from
$\alpha $ to $\gamma $ is the degree of hybridization and hence the
Kondo scale. In a series of publications \cite{Allen} they implemented
this idea mathematically by estimating free energy differences between
these phases by using the solution of a Anderson--Kondo impurity model
supplemented with elastic energy terms. This approach, views the
lattice of Cerium atoms in terms of a single impurity Anderson model,
describing the Ce $f$ electron in fixed bath of band electrons
representing the $spd$ electrons.  The modern Dynamical Mean Field
Theory (DMFT) modifies this simple single impurity picture by
imposing a self--consistency condition on the bath that the $f$
electrons experience. The full power of the DMFT method in combination
with realistic band structure calculations (LDA+DMFT) \cite{LDA+DMFT}
was recently brought to bear on this problem \cite{Zolfl,Held}.
LDA+DMFT allowed the computation of the photoemission spectra of
Cerium in both phases, and the thermodynamics of the transition
starting from first principles. The theoretical results were in good
agreement with existing experiments \cite{expts}. The photoemission
spectra close to the Fermi level, is dominated by the $f$ electron
density of states. The most recent DMFT studies of Cerium demonstrated
that both the Mott transition picture and the Kondo collapse picture
lead to similar photoemission and inverse photoemission spectra.
The spectra of the $\alpha$ phase consists of Hubbard bands and a
quasiparticle peak while the insulating $\gamma$ phase has no
quasiparticle or Kondo peak in the spectra and consists of Hubbard
bands only. Hence, from the study of the $f$ electron spectra which
dominates the photoemission spectroscopy, it is not possible to
decide whether the resonance arises from the proximity to the Mott
transition or from the Kondo effect between the $f$ and the the $spd$
electrons.


In this letter we revisit this problem theoretically using optical
spectroscopy. Our qualitative idea is that the $spd$ electrons have very large
velocities, and therefore they will dominate the optical spectrum of this
material. In the Mott transition picture, the $spd$ electrons are pure
spectators, and hence no appreciable changes in the optical spectrum should
be observed. On the other hand, if the hybridization between the $spd$
electrons and the $f$ electrons increases upon entering the $\alpha$ phase, we
expect a pseudogap to develop as the temperature is lowered because $spd$
carriers are strongly modified as they bind to the $f$ electrons. Optical
experiments on thin films were carried out recently \cite{VanDerMarel}. They
offer a benchmark to compare the results of the theoretical DMFT
calculation. We will show that DMFT reproduces all the features observed in
the experiment. By studying in detail the temperature dependence of the
spectra, we can interpret the different experimental features. We find that
Cerium is much closer to the volume collapse picture than to the Mott
transition picture.

\textit{Formalism} - Within the LDA+DMFT method \cite{LDA+DMFT}, the LDA
Hamiltonian is superposed by a Hubbard--like local Coulomb interaction,
which is the most important source of strong correlations in correlated
materials and is not adequately treated within LDA alone. The resulting
many--body problem is then treated in DMFT spirit, i.e., neglecting the
nonlocal part of self--energy. It is well understood by now, that this theory
is not only exact in the limit of infinite dimensions but is also a very
valuable approximation for many three dimensional systems since it is
capable of treating delocalized electrons from LDA bands as well as
localized electrons on equal footing.

To calculate the central object of DMFT, the local Green's function
$G_{loc}$, we solved the Dyson equation
\begin{equation}
(H_{{\mathbf{k}}}^{LDA}+\Sigma(\omega )-E_{dc}-\epsilon_{{\mathbf{k}}
j\omega} O_{{\mathbf{k}}})\psi_{{\mathbf{k}}\omega }^{R,j}=0,  \label{Dyson}
\end{equation}
where $H_{{\mathbf{k}}}^{LDA}$ is the LDA Hamiltonian expressed in a
localized linear muffin--tin orbital (LMTO) base, $O_{{\mathbf{k}}}$
is the overlap matrix appearing due to nonorthogonality of the base
and $\Sigma(\omega)$ is the self--energy matrix to be determined by
the DMFT. A double counting term $E_{dc}$ appears here since the
Coulomb interaction is also treated by LDA in a static way therefore
the LDA local correlation energy has to be subtracted. Given the
eigenvalues $\epsilon_{{\mathbf{k}} j\omega}$, the left eigenvectors
$\psi_{{\mathbf{k}}\omega }^{L,j}$ and the right eigenvectors
$\psi_{{\mathbf{k}}\omega }^{R,j}$ of Eq.~(\ref{Dyson}) the local
Green's function can be expressed by
\begin{equation}
G_{loc,\a\b} = \sum_{{\mathbf{k}} j}\frac{\psi^{R,j}_{{\mathbf{k}}\omega,\a}
\psi^{L,j}_{{\mathbf{k}}\omega,\b}}{\omega+\mu-\epsilon_{{\mathbf{k}}
j\omega}}.  \label{G_loc}
\end{equation}
The local self--energy $\Sigma(\omega)$, that appears in Eq.~(\ref{Dyson}),
can be calculated from the corresponding impurity problem, defined by the
DMFT self--consistency condition
\begin{equation}
G_{loc}=\left( \omega -E_{imp}-\Sigma (\omega )-\Delta _{imp}(\omega
)\right) ^{-1}.  \label{G_imp}
\end{equation}
Here $\Delta _{imp}$ is the impurity hybridization matrix and $E_{imp}$ are
the impurity levels. The solution of the Anderson impurity problem, i.e.,
the functional $\Sigma \lbrack \Delta _{imp}(\omega),E_{imp},U]$, closes the
set of equations (\ref{Dyson}), (\ref{G_loc}) and (\ref{G_imp}).

\begin{figure}[tbp]
\includegraphics[angle=-90,width=0.9\linewidth]{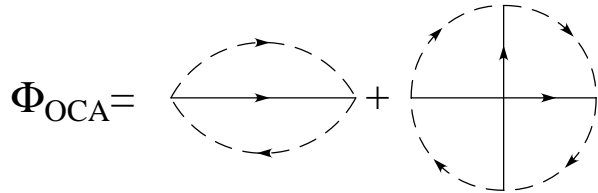}
\caption{ Luttinger--Ward Functional for the impurity solver called
one--crossing approximation (OCA). Full lines correspond to the conduction
electrons from the bath, while dashed lines stand for slave particles. Each
slave particles, however, corresponds to exactly one atomic states. }
\label{Fig1}
\end{figure}

Various many-body techniques can be used to solve the impurity
problem, among others, the Quantum Monte Carlo method (QMC),
Non--crossing approximation (NCA) or iterative perturbation theory
(IPT). Here, we used One--crossing approximation (OCA)
\cite{Pruschke,Haule}, which is, like NCA, a self-consistent
diagrammatic method that perturbes on atomic limit. Unlike NCA, it
takes into account certain type of vertex correction, namely all
diagrams that do not contain a line with more than one crossing, hence
the name, One crossing approximation. This approximation is far
superior to NCA, as is well known, considerably improves the
Fermi-liquid scale $T_K$ and reduces some pathologies of NCA, but is
more time consuming. The OCA is also the lowest order self-consistent
approximation exact up to $V^2\propto \Delta_{imp}$ where
$\Delta_{imp}$ is the impurity hybridization with the electronic
bath. A self-consistent diagrammatic approximation is most
conveniently expressed by its Luttinger-Ward functional $\Phi$, which
uniquely determines the dressing of the slave particles, i.e.,
$\Sigma_{\alpha\beta}={\partial\Phi/\partial G_{\beta\alpha}}$. First
diagram of $\Phi$ in Fig.~\ref{Fig1} corresponds to the well known
NCA, while the second, i.e., the crossing diagram, defines corrections
that are taken into account in OCA.

The optical conductivity is given by a current--current correlation function
which can be expressed in the following way
\begin{widetext}
\begin{eqnarray}
\sigma _{\mu \nu }(\omega ) =\frac{e^2}{2\pi}\sum_{ss^\prime=\pm}
ss^{\prime}\sum_{\vk jj^\prime}\int
d\varepsilon\frac{f(\varepsilon^+)-f(\varepsilon^-)}{\omega}
\frac{M_{\vk jj^\prime}^{ss^\prime,\,\mu\nu}(\varepsilon^-,\varepsilon^+)}
     {\omega+\epsilon_{\vk j\varepsilon^-}^s-\epsilon_{\vk
     j^\prime\varepsilon^+}^{s^\prime}}
\left[\frac{1}{\varepsilon ^{-}+\mu -\epsilon _{\vk j\varepsilon
^{-}}^{s}}{-}\frac{1}{\varepsilon ^{+}+\mu -\epsilon _{\vk j^{\prime
}\varepsilon ^{+}}^{s^{\prime }}}\right] ,  \label{OPT}
\end{eqnarray}
\end{widetext}where we have denoted $\varepsilon ^{\pm }=\varepsilon \pm
\omega /2$, and used the shortcut notations $\epsilon _{{\mathbf{k}}
j\varepsilon }^{+}\equiv \epsilon _{{\mathbf{k}}j\varepsilon }$, $\epsilon
_{ {\mathbf{k}}j\varepsilon }^{-}\equiv\epsilon _{{\mathbf{k}}j\varepsilon
}^{\ast }$. The matrix elements $M_{{\mathbf{k}}jj^{\prime }}$ appear as
standard dipole allowed transition probabilities which are now defined with
the right and left solutions $\psi^{R}$ and $\psi^{L}$ of the Dyson
equation:
\begin{eqnarray}
&&M_{{\mathbf{k}} jj^\prime}^{ss^\prime,\,\mu\nu}(\varepsilon^-,
  \varepsilon^+)= \sum_{\a_1,\a_2}
  (\psi_{{\mathbf{k}}\varepsilon^-,\a_1}^{j, s})^s
  v_{{\mathbf{k}}\a_1\a_2}^{\mu} (\psi_{{\mathbf{k}}\varepsilon^+,\a
  _2}^{j^\prime, -s^\prime})^{s^\prime}\times \notag \\
  &&\qquad\qquad\qquad\qquad\sum_{\a_3,\a_4}
  (\psi_{{\mathbf{k}}\varepsilon^+,
  \a_3}^{j^\prime,s^\prime})^{s^\prime}v_{{\mathbf{k}}\a_3\a_4}^{\nu}
  (\psi_{{\mathbf{k}}\varepsilon^-,\a_4}^{j,-s})^s, \label{MAT}
\end{eqnarray}
where we have denoted
$\psi_{{\mathbf{k}}\varepsilon}^{j,+}\equiv\psi_{{\mathbf{k}}\varepsilon}^{j,L}$
, $\psi_{{\mathbf{k}}\varepsilon}^{j,-}\equiv
\psi_{{\mathbf{k}}\varepsilon}^{j,R}$ and assumed that
$(\psi_{{\mathbf{k}}
\varepsilon}^{j,s})^{+}\equiv\psi_{{\mathbf{k}}\varepsilon}^{s,j}$
while $
(\psi_{{\mathbf{k}}\varepsilon}^{j,s})^{-}\equiv(\psi_{{\mathbf{k}}\varepsilon}^{j,s})^\ast$.
The expressions (\ref{OPT}),(\ref{MAT}) represent generalizations of
the formulae for optical conductivity for a strongly correlated
system, and involve the extra internal frequency integral appearing in
Eq.~(\ref{OPT}).

As in previous work \cite{Zolfl,Held} we treat only the $f$--electrons
as strongly correlated thus requiring full energy resolution, while
all other electrons such as Ce $s,p,d$ are assumed to be well
described by the LDA.  The spin--orbit coupling is fully taken into
account therefore the size of the matrixes in Eq.~(\ref{Dyson}) is
$32\times 32$, while the self--energy matrix appears in a sub--block
of size $14\times 14$. The value of the Coulomb interaction $U$ was
calculated by the local density constrained occupation calculations
\cite{McMahan} and found to be $U\sim 6$ eV. After the
self--consistency of the dynamical mean field equations is reached,
the quasiparticle spectra described by $\psi_{{\mathbf{k}}\omega}^j$
and $\epsilon_{{\mathbf{k}} j\omega }$ are found, and the expression
(\ref{OPT}) for $\sigma_{\mu\nu}(\omega)$ can be evaluated. Here, we
paid a special attention to the energy denominator $1/(\omega
+\epsilon _{{\mathbf{k}} j\varepsilon ^{-}}^{s}-\epsilon
_{{\mathbf{k}}j^{\prime }\varepsilon ^{+}}^{s^{\prime }})$ appearing
in (\ref{OPT}). Due to its strong ${\mathbf{k}}$--dependence the
tetrahedron method of Lambin and Vigneron \cite{Lambin} is used. The
integral over internal energy $\varepsilon $ is calculated in an
analogous way. The frequency axis is divided into discrete set of
points $\varepsilon _{i}$ and assumed that the eigenvalues $\epsilon
_{{\mathbf{k}} j\varepsilon }$ and the matrix elements
$M_{{\mathbf{k}}jj^{\prime }}^{ss^{\prime },\,\mu \nu }(\varepsilon
^{-},\varepsilon ^{+})$ can be linearly interpolated between each pair
of points. We first perform the frequency integration to convert the
double pole expression (\ref{OPT}) into a single pole expression for
which the tetrahedron method is best suited.

\begin{figure}[tbp]
\includegraphics[width=0.9\linewidth]{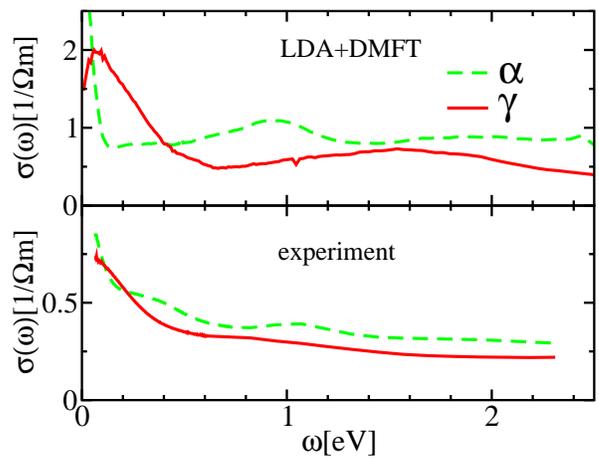}
\caption{ The top panel shows the optical conductivity calculated by LDA+DMFT method for both
$\alpha$ and $\gamma$ phase of Cerium. The volume of
$\alpha$ phase is $28.06{\rm \r{A}}^3$ and the temperature used in
calculation is $580K$.  The volume of $\gamma$ phase is $34.37{\rm
\r{A}}^3$ and $T=1160K$. The bottom panel shows experimental results
measured by the group of D. van der Marel \cite{VanDerMarel}. The
measurements for $\alpha$ phase were done at 5~K and for $\gamma$ phase
at 300~K.
}
\label{Fig2}
\end{figure}
\begin{figure}[tbp]
\includegraphics[width=0.9\linewidth]{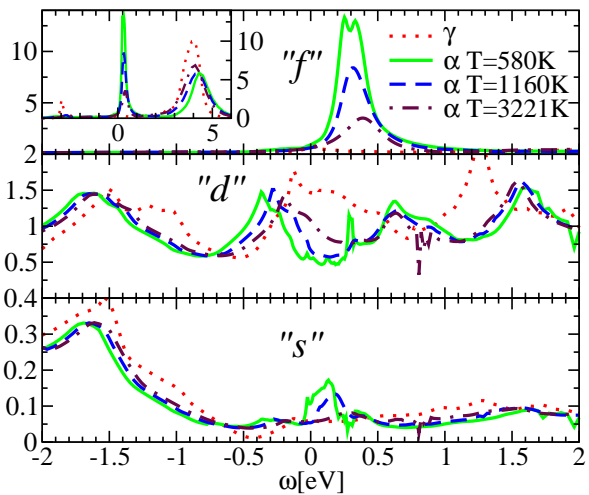}
\caption{ Temperature dependence of partial density of states for both
phases $\protect\alpha $ and $\protect\gamma $ of Ce. }
\label{Fig5}
\end{figure}

\textit{Results} - What distinguishes the $\alpha$ phase from the
$\gamma$ phase when looking at the optical response of Cerium? How do
the calculated results compare with experiments? In the top panel of
Fig.~\ref{Fig2} we present the calculation of LDA+DMFT method while in
the bottom panel the results of experiment on thin films
\cite{VanDerMarel} are reproduced. Notice that in both the experiment
and theory the overall magnitude of conductivity in $\gamma$ phase is
smaller than in $\alpha$ due to the larger volume and hence reduced
velocity of $\gamma$ phase. In the $\alpha$ phase, the Drude peak
width is of the order of $0.1$ eV while in $\gamma $ phase it is at
least an order of magnitude narrower. This provides us with a crude
estimate of $T_{K}$, which is of the order of $1000$ K and $100$ K in
the $\alpha$ and $\gamma $ phase, respectively. The intermediate
frequency range in $\alpha$ is characterized by a dip or pseudogap
ranging from $0.2$ eV to $0.8 $ eV and a peak around $1$ eV. This gap
is filled--in in $\gamma $ phase resulting in a broad hump up to $0.5$
eV. This nontrivial behavior can be understood by examining the
orbitally resolved partial density of states displayed in Fig.~\ref{Fig5}.

The largest contribution to the total density of states around the
Fermi level comes from the $f$ bands which have much smaller
velocities and, at the same time, almost all spectral weight above the
Fermi level. Hence, the transitions from below to above the Fermi
level have a small amplitude therefore $f$ bands make a very small
contribution to the optical response.  Indeed, we found that the
contribution involving only $f$ electrons is orders of magnitude
smaller than contribution of the $d$ bands. Since $s$ and especially
$p$ bands have very little spectral weight close to the Fermi level,
the most important contribution to the optical response comes from the
$d$ bands.

\begin{figure}[tbp]
\includegraphics[width=0.8\linewidth]{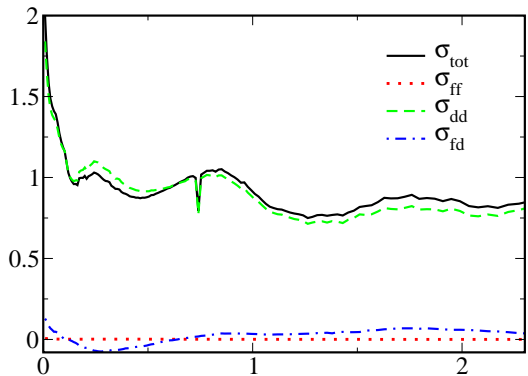}
\caption{ The partial contribution to the optical conductivity of
$\protect\alpha$ phase of $Ce$ at $T=1160K$. }
\label{FigN}
\end{figure}
To shed some light on the states probed optically  we introduce
an orbital  resolution of the optical conductivity (or "fat
optics").
\begin{equation}
\sigma(\omega) = \sigma^{ff}(\omega)+\sigma^{dd}(\omega)+\sigma^{fd}(\omega)
\end{equation}
where $ff$ stands for the pure $f$ contribution, $dd$ for pure $spd$
contribution and $fd$ for the mixed part. The partial conductivities
are calculated by taking the appropriate dipole matrix elements
$M_{{\mathbf{k}}jj^{\prime}}$ in Eq.~(\ref{OPT}), for example, to
calculate $ff$ part all four indexes $\a_1,\a_2,\a_3,\a_4$ in
Eq.~(\ref{MAT}) run over $f$ orbitals only while in $dd$ part all
indexes run over $spd$ orbitals. The results displayed in
Fig.~\ref{FigN}, demonstrate that indeed the optical conductivity has
very little $f$ character.

The small contribution of the $f$ electrons to the  optical
conductivity, however, does not imply that the large Kondo peak,
which has $f$ character, is irrelevant for the optical
conductivity. The $f$ electrons, mixing with the $spd$ bands,
form a large Kondo peak that increases as the temperature is
reduced and eventually induces a hybridization pseudogap in $spd$
spectra. As one can see in Fig.~\ref{Fig5}, the $d$ bands in
$\alpha$ phase have a very pronounced pseudogap which is growing
by lowering temperature exactly as the Kondo peak builds up. The
spectral weight is transferred from the Fermi level into the
side--peaks which are $1$ eV apart causing $1$ eV peak in optical
conductivity. In the $\gamma$ phase, the Kondo peak disappears
because the effective hybridization of the $f$ with $spd$
electrons is smaller and Kondo scale $T_{K}$ is reduced for at
least an order of magnitude. In this regime the hybridization
pseudogap in the $d$ bands is not formed, and instead we find a
broad hump in the optical response.

Although all basic features of measured optical response are well
explained by LDA+DMFT calculation, there are some discrepancies.  The
overall magnitude of measured conductivity in both phases is
approximately half of the calculated value. This discrepancy might be
a consequence of negligence of interactions on the $spd$ electrons, an
effect which would reduce the current matrix elements. Early
measurements by Rhee. et.al. \cite{Rhee}, hovewer, suggest that the
optical conductivity is indeed for factor of two larger than the one
measured by the group of D. van der Marel \cite{VanDerMarel}. Also
the shoulder in $\alpha $ phase, appearing around $0.33$ eV in
Fig.~\ref{Fig2}b, is absent in LDA+DMFT results.

\begin{figure}[tbp]
\includegraphics[width=0.9\linewidth]{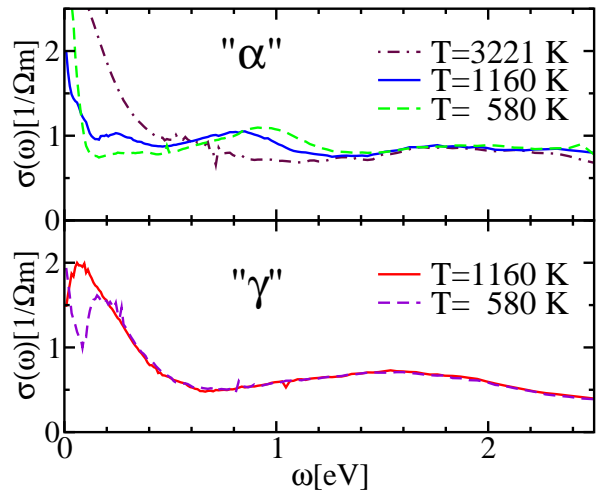}
\caption{
  Temperature dependence of optical conductivity for both
  phases $\protect\alpha $ and $\protect\gamma $ of Ce.  }
\label{Fig4}
\end{figure}
The temperature dependence of the optical response for both the $\alpha$ and
$\gamma$ phase are shown in Fig.~\ref{Fig4}. With increasing temperature,
the $1$ eV peak moves to smaller frequencies and finally at few $1000$ K
disappears. At the same time, the pseudogap below the peak gradually
disappears as the temperature is raised and evolves into a broad hump, just
like the one in $\gamma$ phase. This temperature dependence is easily
understood by looking at the partial density of states in Fig.~\ref{Fig5}.
Since the Kondo peak is gradually reduced with increasing temperature the
hybridization pseudogap in $d$ bands disappears causing featureless optical
response. In $\gamma$ phase, the Kondo peak is very small and causes some
temperature dependence at very low temperatures only at very low frequency
below $0.1$ eV.

\textit{Conclusion} - The methodology we introduced, allow us to
interpret the optical spectra of $\alpha$ and $\gamma$ Cerium, in favor
of the Kondo volume collapse model. The formalism should be useful
in many other problems of correlated electrons where temperature
dependent transfer of spectral weight between widely different
energy ranges takes place, a phenomena which can be beyond the
scope of simple model Hamiltonians.

\end{document}